\documentclass[letterpaper, 10 pt, conference]{ieeeconf}
\IEEEoverridecommandlockouts
\usepackage[T1]{fontenc}
\usepackage{amsmath,amssymb,amsthm,bm,mathrsfs}
\usepackage{algorithm,float}
\usepackage{booktabs,graphicx,xcolor,cite,url,balance}
\newtheorem{assumption}{Assumption}
\newtheorem{lemma}{Lemma}
\newtheorem{theorem}{Theorem}
\newtheorem{remark}{Remark}
\DeclareMathOperator{\tr}{tr}
\DeclareMathOperator{\sym}{sym}
\DeclareMathOperator{\off}{off}
\newcommand{\E}{\mathbb{E}}

\newcommand{\R}{\mathbb{R}}

\title{\LARGE \bf Estimating Hessian Log-Determinants\\without Forming the Hessian Matrix}
\author{Geng Zhang$^{1}$ and James C. Spall$^{2}$%
\thanks{$^{1}$G. Zhang  is a graduate student of the Johns Hopkins University Applied Math and Statistics Department. 3400 North Charles Street, Baltimore, MD 21218 {\tt\small gzhang37@jhu.edu}}%
\thanks{$^{2}$James C. Spall is a member of the Principal Professional Staff at the JHU Applied Physics Laboratory and a Research Professor of the Johns Hopkins University Applied Math and Statistics Department. 3400 North Charles Street, Baltimore, MD 21218 {\tt\small James.Spall@jhuapl.edu}}%
}

\begin{document}
\maketitle
\thispagestyle{empty}
\pagestyle{empty}

\begin{abstract}
The log-determinant of a positive-definite Hessian matrix is used in uncertainty assessment, Bayesian model comparison, and experimental design. Computing this quantity can be difficult when function values are available but derivatives are not. We combine simultaneous perturbation stochastic approximation with stochastic Lanczos quadrature (SPSA-SLQ) to estimate the log-determinant without forming the Hessian. Possibly noisy function values suffice for fixed-data and expected Hessians, including Fisher information matrices. We derive a local bound on log-determinant error due to Hessian-vector product approximation, quadrature, and averaging random probes. Three numerical studies examine accuracy and computational cost. In a 3002-parameter example, SPSA-SLQ gives a relative root-mean-square error below 1\% and uses 65\% less peak GPU memory than a basic full-matrix estimator. A sensor-placement study also compares improved full-matrix methods and evaluates layout selection.
\end{abstract}

\section{Introduction}
\label{sec:introduction}

Mathematical models support prediction and control. After fitting a model, one may assess parameter uncertainty, compare models, or choose additional measurements \cite{ljung1999}. The Hessian describes local criterion curvature, and its determinant enters uncertainty and model-comparison calculations. In this paper, we develop an efficient method for estimating a positive-definite Hessian log-determinant without calculating or estimating all Hessian entries.

The Fisher information matrix (FIM) describes the information observations provide about model parameters \cite{spall2005}. Under a large-sample Gaussian approximation, its determinant determines a parameter confidence-ellipsoid volume. Maximizing it gives D-optimal design \cite{pronzato2008}, used in sensor selection \cite{joshi2009}, mobile-sensor trajectories \cite{tricaud2008acc}, and battery input design \cite{park2018acc}. Information determinants also enter robot-motion analysis for angle-of-arrival estimation \cite{jadhav2022}.

Beyond Fisher information, Bayesian model comparison uses a negative log-posterior Hessian determinant in a Laplace approximation to model evidence \cite{beck2004}, with related calculations in dynamic system identification \cite{zhou2022}. A local probabilistic model for inverse optimal control involves the Hessian of negative total reward with respect to the control sequence \cite{levine2012}.

Cholesky factorization of a dense $d\times d$ positive-definite Hessian uses $\mathcal O(d^3)$ operations and $\mathcal O(d^2)$ storage, excluding the cost of obtaining its entries. Simulation-based identification may provide objective values while derivatives are expensive or unavailable \cite{gedda2012}. We consider scalar objectives with continuous derivatives and possibly noisy evaluations, including objectives computed from vector-valued system outputs.

Simultaneous perturbation stochastic approximation (SPSA) estimates gradient vectors by perturbing all parameters together, using a fixed number of generally noisy function evaluations per replicate \cite{spall1992,spall2000}. SPSA-based full information-matrix estimators \cite{spall2005} include variance-reduction improvements \cite{spall2008acc}. Randomized trace estimation \cite{hutchinson1989} averages quadratic forms in $\log\bm H$ to estimate its trace; stochastic Lanczos quadrature (SLQ) \cite{golubmeurant2010,ubaru2017} approximates these forms using Hessian-vector products and small matrices.

We derive a local log-determinant error bound and use three information-matrix studies to compare accuracy, cost, and sensor-layout selection. Other applications motivate the general formulation.

\section{Problem Formulation}
\label{sec:problem}

Let $\Phi:\R^d\to\R$ be a scalar criterion, such as a negative log-likelihood or negative log-posterior density, with $d$ adjustable parameters $\bm\theta$. Assume continuous partial derivatives through order four near a specified point $\bm\theta_0$ and a positive-definite Hessian there. Our aim is to estimate
\begin{equation}
\begin{gathered}
\mathcal L=\log\det\bm H=\tr(\log\bm H),\\
\text{where}\quad \bm H=\nabla^2\Phi(\bm\theta_0)\succ\bm0.
\end{gathered}
\label{eq:target}
\end{equation}
The identity in \eqref{eq:target} appears in \cite{ubaru2017}. Here $H_{ij}=\partial^2\Phi/\partial\theta_i\partial\theta_j$ at $\bm\theta_0$, and $\tr$ sums diagonal entries. Global convexity of $\Phi$ is not required. All scalar logarithms are natural. Write the eigendecomposition $\bm H=\bm U\operatorname{diag}(\lambda_1,\ldots,\lambda_d)\bm U^\top$, with orthogonal $\bm U$ and $\lambda_i>0$. Then its matrix logarithm is $\log\bm H=\bm U\operatorname{diag}(\log\lambda_1,\ldots,\log\lambda_d)\bm U^\top$. Both $\log\det\bm H$ and $\tr(\log\bm H)$ equal $\sum_{i=1}^d\log\lambda_i$. We use $\|\cdot\|_2$ for the Euclidean vector and induced matrix norms, $\|\cdot\|_F$ for the Frobenius norm, and $\bm I_d$ for the identity matrix.

A routine evaluates $\Phi$ at chosen parameter inputs, possibly with noise:
\begin{equation}
\widetilde\Phi(\bm x;\xi)=\Phi(\bm x)+\epsilon(\bm x;\xi).
\label{eq:oracle}
\end{equation}
The random input $\xi$ may represent simulated data, and $\epsilon=0$ for noise-free evaluations. For an expected criterion, $\phi(\bm\theta;\xi)$ is the scalar loss for one realization of $\xi$: $\Phi(\bm\theta)=\E_\xi[\phi(\bm\theta;\xi)]$ and $\widetilde\Phi(\bm\theta;\xi)=\phi(\bm\theta;\xi)$. With the distribution of $\xi$ fixed and with differentiation under the expectation permitted, $\bm H=\E_\xi[\nabla^2\phi(\bm\theta_0;\xi)]$. We estimate $\log\det\bm H$, not the mean of sample log-determinants. Available derivatives can also be used (Section~\ref{sec:estimator}).

\emph{Fisher information.} Let $\bm Y$ contain $n$ possibly vector-valued observations with joint density or probability mass function $p(\bm Y\,|\,\bm\theta)$. With score $\bm s(\bm Y;\bm\theta)=\nabla_{\bm\theta}\log p(\bm Y\,|\,\bm\theta)$, the standard FIM is $\bm F_n(\bm\theta_0)=\E_{\bm\theta_0}[\bm s(\bm Y;\bm\theta_0)\bm s(\bm Y;\bm\theta_0)^\top]$ \cite{spall2005}. To express this matrix as the Hessian in \eqref{eq:target}, take $\Phi=\Phi_{\bm\theta_0}$, the expected negative log-likelihood under a fixed generating distribution \cite{white1982}:
\begin{equation}
\Phi_{\bm\theta_0}(\bm\theta)=\E_{\bm Y\sim p(\cdot\,|\,\bm\theta_0)}[-\log p(\bm Y\,|\,\bm\theta)].
\label{eq:crossentropy}
\end{equation}
The nominal parameter $\bm\theta_0$ may be an estimate or design value. It remains fixed in the expectation while $\bm\theta$ varies in the likelihood. For a regular model with parameter-independent support and differentiation under the expectation permitted,
\begin{equation}
\begin{aligned}
\left.\nabla^2_{\bm\theta}\Phi_{\bm\theta_0}(\bm\theta)\right|_{\bm\theta=\bm\theta_0}
&=-\E_{\bm\theta_0}\!\left[\left.\nabla^2_{\bm\theta}\log p(\bm Y\,|\,\bm\theta)\right|_{\bm\theta=\bm\theta_0}\right]\\
&=\bm F_n(\bm\theta_0).
\end{aligned}
\label{eq:fisher-identity}
\end{equation}
The information-matrix identity gives the final equality \cite{spall2005}. In \eqref{eq:oracle}, set $\xi=\bm Y\sim p(\cdot\,|\,\bm\theta_0)$ and $\widetilde\Phi(\bm\theta;\bm Y)=-\log p(\bm Y\,|\,\bm\theta)$. Even if computed exactly for realized data, this value is random relative to \eqref{eq:crossentropy}, with $\epsilon=\widetilde\Phi-\Phi_{\bm\theta_0}$. Fixed-data $\Phi(\bm\theta)=-\log p(\bm y\,|\,\bm\theta)$ instead gives observed information. Both fit \eqref{eq:target} when positive definite, but generally differ.

For a $(1-\alpha)$ probability level, $0<\alpha<1$, the large-sample normal approximation with covariance $\bm F_n(\bm\theta_0)^{-1}$ gives an approximate confidence ellipsoid in $\R^d$: $(\bm u-\widehat{\bm\theta})^\top\bm F_n(\bm\theta_0)(\bm u-\widehat{\bm\theta})\le q_\alpha$, centered at the parameter estimate $\widehat{\bm\theta}$. Its volume is $V=v_d q_\alpha^{d/2}/\sqrt{\det\bm F_n(\bm\theta_0)}$, where $v_d$ is the unit-ball volume and $q_\alpha$ is the $(1-\alpha)$ quantile of $\chi_d^2$; both depend on $d$. Thus $\log V=C-\tfrac12\log\det\bm F_n(\bm\theta_0)$, with $C$ fixed for the same dimension, confidence level, and parameter units \cite{pronzato2008}. D-optimal placement minimizes this approximate volume (Example~3).

\emph{Bayesian model evidence.} For fixed data $\bm y$ and a normalized prior density $\pi$, choose $\Phi(\bm\theta)=-\log p(\bm y\,|\,\bm\theta)-\log\pi(\bm\theta)$. At a posterior mode $\widehat{\bm\theta}$ in the interior of the parameter domain, with $\nabla^2\Phi(\widehat{\bm\theta})\succ\bm0$, set $\bm\theta_0=\widehat{\bm\theta}$. The Laplace approximation to the marginal likelihood $p(\bm y)$ gives
\begin{equation}
\log p(\bm y)\approx-\Phi(\widehat{\bm\theta})+\tfrac d2\log(2\pi)-\tfrac12\log\det\bm H.
\label{eq:laplace}
\end{equation}
Here $\bm H$ includes prior curvature and generally differs from the FIM \cite{beck2004,zhou2022}. Likelihood and prior evaluations supply function values for our estimator; the mode and Laplace approximation are taken as given.

\section{Matrix-Free SPSA-SLQ Estimator}
\label{sec:estimator}

A Hessian-vector product (HVP), $\bm H\bm v\in\R^d$, is the directional derivative of the gradient vector $\nabla\Phi$ at $\bm\theta_0$ in a direction $\bm v$. We estimate HVPs from function values using SPSA (Subsection~\ref{subsec:spsahvp}), then use small matrices to approximate and average $\bm z^\top\log(\bm H)\bm z$ (Subsection~\ref{subsec:projection}).

\subsection{SPSA Hessian-Vector Products}
\label{subsec:spsahvp}

We estimate gradients at $\bm\theta_0\pm h\bm v$, with a direction $\bm v$ satisfying $\|\bm v\|_2=1$, directional step $h>0$, and SPSA step $c>0$. The perturbation $\bm\Delta_k=(\Delta_{k1},\ldots,\Delta_{kd})^\top$ has independent entries taking $+1$ or $-1$, each with probability $1/2$. With $c$ fixed within each HVP estimate, let $y_k^{(\pm)}(\bm x)=\widetilde\Phi(\bm x\pm c\bm\Delta_k;\xi_k^{(\pm)}(\bm x))$. These are possibly noisy criterion measurements in \eqref{eq:oracle}; for the FIM, they are negative log-likelihood values from simulated data. The $i$th component of the SPSA gradient estimate \cite{spall1992} is
\begin{equation}
\widehat g_{ki}(\bm x)=\frac{y_k^{(+)}(\bm x)-y_k^{(-)}(\bm x)}{2c\Delta_{ki}},
\qquad i=1,\ldots,d.
\label{eq:spsa-gradient}
\end{equation}
Averaging $N$ independent central differences reduces variance in estimating $\bm H\bm v$. Using the same $\bm\Delta_k$ at both points gives
\begin{equation}
\widehat{\bm H\bm v}_N=\frac1N\sum_{k=1}^{N}\frac{\widehat{\bm g}_k(\bm\theta_0+h\bm v)-\widehat{\bm g}_k(\bm\theta_0-h\bm v)}{2h}.
\label{eq:hvp}
\end{equation}
Each of the $N$ independent estimates of $\bm H\bm v$ uses a fresh $\bm\Delta_k$ and four function evaluations at $\bm\theta_0\pm h\bm v\pm c\bm\Delta_k$. Random inputs may be shared within a replicate. For the FIM, pseudodata are simulated samples $\bm Y_k\sim p(\cdot\,|\,\bm\theta_0)$: evaluate each sample's negative log-likelihood at all four points and draw independently for the next replicate from the same fixed distribution, as in \eqref{eq:crossentropy}.

For a known quadratic term $\Phi(\bm x)=\phi(\bm x)+\lambda\|\bm x-\bm\theta_0\|_2^2/2$ with regularization weight $\lambda\ge0$, add $\lambda\bm v$ exactly to the SPSA product for $\phi$. Example~1 uses this decomposition with $\nabla^2\phi(\bm\theta_0)\succeq\bm0$.

\subsection{Orthogonalized Projection}
\label{subsec:projection}

A probe is a Rademacher vector $\bm z$ whose entries independently take $\pm1$ with equal probability. The identity $\E[\bm z^\top\log(\bm H)\bm z]=\log\det\bm H$ motivates averaging quadratic forms. Each is approximated by $d\bm e_1^\top\log(\bm S_m)\bm e_1$, where $\bm e_1$ is the first coordinate vector in $\R^m$ and $\bm S_m$ is the $m\times m$ matrix constructed below. Only the smaller matrix logarithm is evaluated.

Choose projection depth $1\le m\le d$ and probe count $n_v\ge1$. Draw independent probes $\bm z^{(\ell)}\in\{-1,1\}^d$, independently of the fresh SPSA perturbations and random draws driving the function evaluations, and set $\bm q_1^{(\ell)}=\bm z^{(\ell)}/\sqrt d$. For a fixed probe, suppress $\ell$. The unit vectors $\bm q_j$ are generated from $\bm q_1$ by \eqref{eq:orthog}. At step $j$, set $\bm v=\bm q_j$ in \eqref{eq:hvp} and average $N$ independent four-evaluation estimates of $\bm H\bm q_j$ to obtain
\begin{equation}
\widehat{\bm u}_j=\widehat{\bm H\bm q_j}_N=\bm H\bm q_j+\bm\omega_j,
\label{eq:inexact-product}
\end{equation}
where $\bm\omega_j$ is the HVP error. With $\bm Q_j=[\bm q_1,\ldots,\bm q_j]\in\R^{d\times j}$, remove the components along its columns and normalize the remaining vector:
\begin{equation}
\bm r_j=(\bm I_d-\bm Q_j\bm Q_j^\top)\widehat{\bm u}_j,
\qquad
\bm q_{j+1}=\bm r_j/\|\bm r_j\|_2\quad(j<m).
\label{eq:orthog}
\end{equation}
For nonzero residuals, $\bm Q_m$ has orthonormal columns spanning the generated subspace in exact arithmetic. We use two orthogonalization passes. After $m$ products, define $\widehat{\bm U}_m=[\widehat{\bm u}_1,\ldots,\widehat{\bm u}_m]$ and $\bm W_m=[\bm\omega_1,\ldots,\bm\omega_m]$. For $\sym(\bm A)=(\bm A+\bm A^\top)/2$, define the $m\times m$ symmetric matrix $\bm S_m$, which approximates $\bm H$ on this subspace, by
\begin{align}
\bm S_m&=\sym(\bm Q_m^\top\widehat{\bm U}_m)\nonumber\\
&=\bm Q_m^\top\bm H\bm Q_m+\sym(\bm Q_m^\top\bm W_m).
\label{eq:reduced-decomposition}
\end{align}
The terms are the projected Hessian and the symmetrized HVP error.

Restoring $\ell$, $\bm S_m^{(\ell)}$ denotes the matrix \eqref{eq:reduced-decomposition} constructed from $\bm z^{(\ell)}$. When all $n_v$ projections complete with positive-definite matrices, the log-determinant estimate is
\begin{equation}
\widehat{\mathcal L}=\frac{d}{n_v}\sum_{\ell=1}^{n_v}\bm e_1^\top\log(\bm S_m^{(\ell)})\bm e_1,
\label{eq:estimator}
\end{equation}
Evaluate the quadratic form using the eigendecomposition of $\bm S_m^{(\ell)}$, not its log-determinant. With exact HVPs, full orthogonalization gives symmetric Lanczos before exact termination, hence standard SLQ.

Fresh stochastic HVPs need not correspond to one symmetric matrix; $\bm S_m$ is therefore symmetrized and generally dense. We compare with exact-product Lanczos \cite{simoncini2003,chen2016}. Available exact or unbiased stochastic gradients can replace SPSA in \eqref{eq:hvp}; available HVPs enter the projection directly. These alternatives require corresponding error bounds; our SPSA analysis and evaluation counts concern function values.

Algorithm~\ref{alg:estimator} uses residual tolerance $\tau=10^{-12}$ to avoid division by a nearly zero norm and checks positivity of the $m\times m$ matrix. Either failed check stops the attempt without an estimate; nonpositive eigenvalues are not clipped.

\begin{algorithm}[t]
\caption{SPSA-SLQ from function evaluations}
\label{alg:estimator}
\textbf{Given:} The evaluation routine \eqref{eq:oracle} and point $\bm\theta_0$ in \eqref{eq:target}; depth $m$ and probe count $n_v$ (Section~\ref{subsec:projection}); replication count $N$ and steps $c,h$ in \eqref{eq:spsa-gradient}--\eqref{eq:hvp}; residual tolerance $\tau>0$.

Perform Steps 1--3 independently for each $\ell=1,\ldots,n_v$, followed by Step 4.
\begin{enumerate}
\item \textbf{Start a projection.} Draw the entries of $\bm z^{(\ell)}$ independently as $+1$ or $-1$ with probability $1/2$. Set $\bm q_1=\bm z^{(\ell)}/\sqrt d$.
\item \textbf{Estimate products and build the basis.} For $j=1,\ldots,m$, use \eqref{eq:hvp} with $\bm v=\bm q_j$ to average $N$ independent four-evaluation estimates of $\bm H\bm q_j$, denoting the result by $\widehat{\bm u}_j$. For $j<m$, form $\bm r_j$ by \eqref{eq:orthog} and either stop without an estimate if $\|\bm r_j\|_2\le\tau$ or normalize it to obtain $\bm q_{j+1}$ otherwise.
\item \textbf{Evaluate the probe contribution.} Form $\bm S_m^{(\ell)}$ by \eqref{eq:reduced-decomposition}. If any eigenvalue is nonpositive, stop without an estimate. Otherwise compute the scalar $d\bm e_1^\top\log(\bm S_m^{(\ell)})\bm e_1$ from its eigendecomposition.
\item \textbf{Average over probes.} Return the arithmetic mean of the $n_v$ scalar contributions, namely $\widehat{\mathcal L}$ in \eqref{eq:estimator}.
\end{enumerate}
\end{algorithm}

There are two averages: $N$ four-evaluation replications produce one HVP estimate, and $n_v$ probe contributions produce $\widehat{\mathcal L}$. With $m$ products per probe, a completed run therefore uses $4n_vmN$ function evaluations; an interrupted run uses fewer.

\section{Finite-Depth Log-Determinant Error Analysis}
\label{sec:error}

We analyze the error in estimating $\log\det\bm H$, where $\bm H=\nabla^2\Phi(\bm\theta_0)$, in exact arithmetic. Theorem~\ref{thm:main} combines HVP and projection, quadrature, and trace-sampling errors.

\subsection{Conditional HVP Bias and Mean-Square Error}

Let $\mathscr F$ contain the trace probes and information available before an HVP call. Conditioning on $\mathscr F$ fixes its adaptive unit direction $\bm v$.

\begin{assumption}
\label{ass:oracle}
On a convex ball of radius $r_0$ about $\bm\theta_0$, $\Phi$ has continuous derivatives through order four and $\|\nabla^4\Phi\|_{\mathrm{op}}\le M_4$, where the tensor norm is the supremum over four unit directions. Assume $\mu\bm I_d\preceq\bm H\preceq L\bm I_d$, $0<\mu\le L$, and $h+c\sqrt d<r_0$; $\mu,L$ bound the Hessian eigenvalues. Conditional on $\mathscr F$, the $N$ replications are independent and identically distributed with fresh Rademacher perturbations. Writing $\epsilon_k^{(\pm)}(\bm x)$ for the error in $y_k^{(\pm)}(\bm x)$, define
\begin{align}
\zeta_k(\bm v)={}&\epsilon_k^{(+)}(\bm\theta_0+h\bm v)-\epsilon_k^{(-)}(\bm\theta_0+h\bm v)\nonumber\\
&-\epsilon_k^{(+)}(\bm\theta_0-h\bm v)+\epsilon_k^{(-)}(\bm\theta_0-h\bm v).
\label{eq:contrast}
\end{align}
Uniformly over admissible histories and inputs, assume $\E[\zeta_k\mid\mathscr F,\bm\Delta_k]=0$ and
\begin{equation}
\E\!\left[\left\|\frac{\zeta_k\bm\Delta_k}{4ch}\right\|_2^2\middle|\mathscr F\right]\le\nu_\epsilon(c,h)<\infty.
\label{eq:noise-moment}
\end{equation}
\end{assumption}

Smoothness controls perturbation bias, and $h+c\sqrt d<r_0$ keeps the evaluations local. The noise conditions concern the four-value SPSA contrast. For conditionally independent zero-mean errors with variance at most $\sigma^2$, take $\nu_\epsilon(c,h)=d\sigma^2/(4c^2h^2)$. Shared random inputs are allowed when \eqref{eq:noise-moment} holds directly.

\begin{lemma}[HVP bias and mean-square error]
\label{lem:hvp}
Under Assumption~\ref{ass:oracle}, define
\begin{align}
b_d&=\frac{M_4d}{6}(h^2+dc^2),\label{eq:bias-constant}\\
\delta^2&=b_d^2+\frac{(\sqrt{d-1}L+b_d)^2+\nu_\epsilon(c,h)}{N}.
\label{eq:delta}
\end{align}
For every adaptive unit input $\bm v$,
\begin{align}
\|\E[\widehat{\bm H\bm v}_N-\bm H\bm v\mid\mathscr F]\|_2&\le b_d,\label{eq:conditional-bias}\\
\E[\|\widehat{\bm H\bm v}_N-\bm H\bm v\|_2^2\mid\mathscr F]&\le\delta^2.
\label{eq:conditional-mse}
\end{align}
\end{lemma}
\begin{proof}
The noiseless four-point contrast divided by $4ch$ is
\begin{equation}
D_k=\frac1{4ch}\int_{-h}^{h}\!\int_{-c}^{c}\bm v^\top\nabla^2\Phi(\bm\theta_0+s\bm v+t\bm\Delta_k)\bm\Delta_k\,dt\,ds.
\label{eq:rectangle}
\end{equation}
The first-order Hessian expansion integrates to zero on the centered rectangle; its remainder has norm at most $M_4\|s\bm v+t\bm\Delta_k\|_2^2/2$. Using $\|\bm\Delta_k\|_2^2=d$ gives $D_k\bm\Delta_k=\bm\Delta_k\bm\Delta_k^\top\bm H\bm v+\bm\rho_k$ with $\|\bm\rho_k\|_2\le b_d$. Also,
\begin{equation}
\E\|\bm\Delta_k\bm\Delta_k^\top\bm u-\bm u\|_2^2=(d-1)\|\bm u\|_2^2.
\label{eq:sp-variance}
\end{equation}
The $L^2$ triangle inequality bounds the centered noiseless replication by $\sqrt{d-1}L+b_d$. The noise has zero cross expectation with functions of $\bm\Delta_k$. Add its second moment, divide the variance by $N$, and add the squared bias.
\end{proof}

For a known quadratic term, apply the lemma to $\phi$ and add $\lambda\bm v$ exactly. The same bound holds when $\bm0\preceq\nabla^2\phi(\bm\theta_0)\preceq\bm H$. Equation~\eqref{eq:sp-variance} explains the dimension dependence of SPSA variability; the physical perturbation length is $c\sqrt d$.

\subsection{Projection Stability}

HVP errors change both the reduced matrix and subsequent directions. Normalizing a nearly zero residual can amplify these errors; the next result requires exact residuals bounded away from zero.

For the same probe, let $\overline{\bm Q}_m$ be the exact-HVP basis and $\overline{\bm T}_m=\overline{\bm Q}_m^\top\bm H\overline{\bm Q}_m$ its Lanczos matrix. Fix $\beta_*>2\tau$ and require every exact residual before step $m$ to have norm at least $\beta_*$ (vacuous for $m=1$). Positive definiteness does not ensure this: a probe may lie in an invariant subspace of dimension below $m$.

\begin{lemma}[Local projection perturbation]
\label{lem:projection}
For an exact trajectory satisfying the residual margin, there are constants $C_m,\eta_m>0$, depending only on $m,L,\mu,\beta_*$, such that $e_{\max}=\max_{j\le m}\|\bm\omega_j\|_2\le\eta_m$ ensures completion and
\begin{align}
\|\bm S_m-\overline{\bm T}_m\|_2&\le C_me_{\max},\label{eq:projection-bound}\\
\lambda_{\min}(\bm S_m)&\ge\mu/2,\label{eq:projection-positive}\\
\|\log\bm S_m-\log\overline{\bm T}_m\|_2&\le(2C_m/\mu)e_{\max}.
\label{eq:matrix-log-bound}
\end{align}
\end{lemma}
\begin{proof}
Induction on \eqref{eq:orthog} bounds $\|\bm Q_m-\overline{\bm Q}_m\|_2\le D_m e_{\max}$ for sufficiently small $e_{\max}$, with $D_m$ depending only on $m,L,\beta_*$. Indeed, projection is locally Lipschitz and normalization satisfies $\|\bm a/\|\bm a\|_2-\bm b/\|\bm b\|_2\|_2\le2\|\bm a-\bm b\|_2/\|\bm b\|_2$, with exact residual norm at least $\beta_*$. Choose the error threshold to keep each perturbed residual above $\tau$. Equation~\eqref{eq:reduced-decomposition} then gives $C_m=2LD_m+\sqrt m$. Requiring $C_m\eta_m\le\mu/2$, Weyl's eigenvalue inequality gives positivity. The matrix-logarithm resolvent identity bounds its difference by $\|\bm S_m-\overline{\bm T}_m\|_2\int_0^\infty(\mu/2+t)^{-2}\,dt$, giving the factor $2/\mu$.
\end{proof}

\subsection{Total Log-Determinant Error}

For exact trajectories reaching depth $m$, define the exact trace estimate and its SLQ approximation using the same probes:
\begin{align}
\mathcal L_{\rm tr}&=\frac1{n_v}\sum_{\ell=1}^{n_v}(\bm z^{(\ell)})^\top\log(\bm H)\bm z^{(\ell)},\nonumber\\
\mathcal L_m&=\frac d{n_v}\sum_{\ell=1}^{n_v}\bm e_1^\top\log(\overline{\bm T}_m^{(\ell)})\bm e_1.
\label{eq:exact-comparators}
\end{align}
The final error separates exactly as
\begin{align}
\widehat{\mathcal L}-\log\det\bm H
={}&\underbrace{(\widehat{\mathcal L}-\mathcal L_m)}_{\text{HVP and projection}}+\underbrace{(\mathcal L_m-\mathcal L_{\rm tr})}_{\text{quadrature}}\nonumber\\
&+\underbrace{(\mathcal L_{\rm tr}-\log\det\bm H)}_{\text{trace sampling}}.
\label{eq:total-decomposition}
\end{align}
Let $\Pi_r$ denote polynomials of degree at most $r$ and define the unit-probe quadrature bound
\begin{equation}
E_{\rm G}(m)=2\min_{p\in\Pi_{2m-1}}\max_{x\in[\mu,L]}|\log x-p(x)|.
\label{eq:gauss-error}
\end{equation}
An $m$-step Gauss rule is exact for polynomials through degree $2m-1$. Approximating the logarithm by such a polynomial gives $|\mathcal L_m-\mathcal L_{\rm tr}|\le dE_{\rm G}(m)$ \cite{golubmeurant2010,ubaru2017}. For $\off(\bm A)=\bm A-\operatorname{diag}(\bm A_{11},\ldots,\bm A_{dd})$, Rademacher sampling gives
\begin{equation}
\E\mathcal L_{\rm tr}=\log\det\bm H,\qquad
\operatorname{Var}(\mathcal L_{\rm tr})=\frac2{n_v}\|\off(\log\bm H)\|_F^2,
\label{eq:trace-variance}
\end{equation}
even when $\log\bm H$ is indefinite \cite{hutchinson1989,cortinovis2022}.

\begin{theorem}[Local total log-determinant error]
\label{thm:main}
Under Assumption~\ref{ass:oracle}, consider runs where every probe has exact residual norms at least $\beta_*>2\tau$ before step $m$ and all HVP errors have norm at most $\eta_m$ from Lemma~\ref{lem:projection}. Then
\begin{align}
|\widehat{\mathcal L}-\log\det\bm H|
\le{}&|\mathcal L_{\rm tr}-\log\det\bm H|+dE_{\rm G}(m)\nonumber\\
&+\frac{2dC_m}{\mu n_v}\sum_{\ell=1}^{n_v}\max_{j\le m}\|\bm\omega_j^{(\ell)}\|_2.
\label{eq:total-pathwise}
\end{align}
Define $X_m=\widehat{\mathcal L}-\log\det\bm H$ when these conditions hold for every probe, and $X_m=0$ otherwise. Its restricted root-mean-square error (RMSE) obeys
\begin{align}
\bigl(\E X_m^2\bigr)^{1/2}\le{}&\underbrace{dE_{\rm G}(m)}_{\text{quadrature}}+\underbrace{\sqrt{2/n_v}\|\off(\log\bm H)\|_F}_{\text{trace sampling}}\nonumber\\
&+\underbrace{\frac{2dC_m\sqrt m}{\mu}\delta}_{\text{HVP and projection}}.
\label{eq:total-bound}
\end{align}
\end{theorem}
\begin{proof}
Equation~\eqref{eq:total-decomposition}, Lemma~\ref{lem:projection}, and the quadrature bound give \eqref{eq:total-pathwise}. The squared maximum product error per probe is bounded by the sum over at most $m$ calls, whose expectation is at most $m\delta^2$ by Lemma~\ref{lem:hvp} (set unused errors after stopping to zero). The trace term has unconditional mean zero and variance \eqref{eq:trace-variance}. Setting contributions to zero outside the stated conditions cannot increase their second moments. Minkowski's inequality gives \eqref{eq:total-bound}.
\end{proof}

Equation~\eqref{eq:total-bound} assigns zero error to runs violating the stated conditions. It bounds neither all-run nor conditional RMSE and gives no completion-probability guarantee. For Fisher information set $\bm H=\bm F_n(\bm\theta_0)$; fixed-data posterior criteria also apply. With other terms fixed in \eqref{eq:laplace}, log-determinant error contributes half as much to log evidence; Laplace approximation error is separate.

\subsection{Parameter Dependence and Computational Cost}
\label{subsec:parameters}

\begin{remark}[Dimension and accuracy]
\label{rem:dimension}
For the exact comparators, let the spectral ratio be $\kappa=L/\mu>1$ and $\gamma=(\sqrt\kappa-1)/(\sqrt\kappa+1)$. Chebyshev truncation in \eqref{eq:gauss-error} and centering $\log\bm H$ at $\tfrac12(\log\mu+\log L)\bm I_d$ give
\begin{align*}
dE_{\rm G}(m)&\le\frac{2d\gamma^{2m}}{m(1-\gamma)},\\
\sqrt{2/n_v}\|\off(\log\bm H)\|_F&\le(\log\kappa)\sqrt{\frac{d}{2n_v}}.
\end{align*}
For uniformly bounded $\kappa$ and absolute tolerance $0<\eta_{\rm acc}\le1$, sufficient worst-case scales for these two terms to be of order $\eta_{\rm acc}$ are $m=\mathcal O(1+\log(d/\eta_{\rm acc}))$ and $n_v=\mathcal O(1+d/\eta_{\rm acc}^2)$, subject to admissible $m\le d$ and the exact-trajectory conditions. These are sufficient, not necessary, choices. A per-parameter or relative tolerance must first be converted to the corresponding dimension-dependent absolute tolerance; relative error requires a nonzero reference. HVP error and projection stability still need to be controlled.
\end{remark}

Increasing $N$ reduces HVP variance, not bias $b_d$; both can grow with $d$, requiring adjustment of $N,c,h$. Increasing $n_v$ reduces trace variability; increasing $m$ reduces quadrature error but may amplify product errors through $C_m$. Shared pseudodata can reduce measurement noise, not SPSA variability. Four evaluations per replicate therefore do not imply dimension-independent equal-accuracy cost.

For supplied $\bm H=\bm R^\top\bm R$, Cholesky gives $\log\det\bm H=2\sum_i\log R_{ii}$. Table~\ref{tab:complexity} compares inputs: coordinate finite differences (FD) and full-matrix SPSA \cite{spall2000,spall2005} estimate Hessian entries; standard SLQ uses supplied HVPs \cite{ubaru2017}.

\begin{table}[!htb]
\centering
\caption{Required input and estimator storage at fixed settings. $B=n_vmN$; FD uses one centered coordinate sweep. Probes are processed sequentially. Evaluation and supplied-HVP workspaces are excluded. These are not equal-accuracy costs (Remark~\ref{rem:dimension}).}
\label{tab:complexity}
\small
\setlength{\tabcolsep}{3pt}
\begin{tabular}{@{}lll@{}}
\toprule
Method & Required input & Storage \\
\midrule
Dense Cholesky & Entries of $\bm H$ & $\mathcal O(d^2)$ \\
Coordinate FD + Cholesky & $2d^2+1$ values & $\mathcal O(d^2)$ \\
Full-matrix SPSA & $4B$ values & $\mathcal O(d^2)$ \\
SLQ with supplied HVPs & $n_vm$ products & $\mathcal O(md+m^2)$ \\
SPSA-SLQ & $4B$ values & $\mathcal O(md+m^2)$ \\
\bottomrule
\end{tabular}
\end{table}

Dense Cholesky and spectral correction cost $\mathcal O(d^3)$ operations; Cholesky requires positive definiteness. Full SPSA accumulation costs $\mathcal O(Bd^2)$, versus $\mathcal O(Bd)$ for SPSA-SLQ. Both SLQ rows add $\mathcal O(n_vm^2d)$ orthogonalization and $\mathcal O(n_vm^3)$ reduced-matrix work. Evaluation and supplied-HVP costs are additional. Storage savings require $m\ll d$; storage becomes quadratic when $m$ is proportional to $d$.

\section{Numerical Results}
\label{sec:numerics}

In this section, we give three distinct examples that illustrate the usefulness of SPSA-SLQ. Example~1 uses finite-shot fidelity measurements; Examples~2 and 3 use sample likelihoods as Monte Carlo measurements of an expected criterion. Analytic information supplies references for all three function-value comparisons.

\subsection{Error and Resource Metrics}
\label{subsec:metrics}

The following terms will be used in the three examples here. Let $J$ count fixed model instances and $R$ independent repetitions per instance. For method $a$, let $\widehat{\mathcal L}_{a,s}^{(r)}$ and $\mathcal L_{\star,s}$ denote the estimate and reference. Define empirical root-mean-square error (RMSE) and relative RMSE by
\begin{align}
\operatorname{RMSE}_a&=\left[\frac1{JR}\sum_{s=1}^J\sum_{r=1}^R(\widehat{\mathcal L}_{a,s}^{(r)}-\mathcal L_{\star,s})^2\right]^{1/2},\label{eq:empirical-rmse}\\
\operatorname{rRMSE}_a(\%)&=\frac{100\operatorname{RMSE}_a}{J^{-1}\sum_{s=1}^J|\mathcal L_{\star,s}|}.
\label{eq:normalized-rmse}
\end{align}
These errors use all estimates, without the local restrictions in Theorem~\ref{thm:main}. RMSE is in log-determinant units; rRMSE uses the nonzero mean absolute reference. Examples 1 and 2 use $(J,R)=(3,10)$ and $(1,3)$; Example~3 uses 18 layouts and $R=30$. Times exclude reference calculations. SP denotes simultaneous perturbation. Spectral correction replaces a symmetric matrix's eigenvalues by their absolute values: $|\bm A|=(\bm A^2)^{1/2}$.

\subsection{Example 1: A Simulated Quantum-State Model}
\label{subsec:example1}

This classically simulated example comes from quantum physics. A pure state is a unit complex vector; fidelity is the squared magnitude of the inner product of two states. Quantum Fisher information (QFI) describes state sensitivity, rather than the classical FIM for a specified observation likelihood. Its determinant describes local information volume \cite{xing2020}, with related criteria in quantum control \cite{clouatre2024}. We use product states of two-level systems (qubits) to obtain exact references, adapting the noisy fidelity measurements and baseline of \cite{gacon2021}.

\begin{table}[t]
\centering
\caption{Example 1: rRMSE (\%) at matched fidelity-query budgets, with 30 estimates per method, dimension, and budget. Smaller values are better. Full-QFI is the spectrally corrected baseline \cite{gacon2021}.}
\label{tab:quantum}
\small
\setlength{\tabcolsep}{4pt}
\begin{tabular}{@{}rrrrr@{}}
\toprule
$d$ & $Q/Q_0$ & Full-QFI & SPSA-SLQ & \shortstack{Exact-HVP\\SLQ$^\dagger$} \\
\midrule
32 & 1 & 6.64 & 8.29 & 7.57 \\
32 & 2 & 4.23 & 8.32 & 7.57 \\
32 & 4 & 2.74 & 5.79 & 5.01 \\
\addlinespace
64 & 1 & 39.95 & 8.15 & 6.13 \\
64 & 2 & 20.13 & 7.64 & 6.13 \\
64 & 4 & 10.18 & 6.32 & 4.83 \\
\addlinespace
128 & 1 & 28.80 & 10.11 & 4.14 \\
128 & 2 & 57.20 & 6.21 & 4.14 \\
128 & 4 & 52.32 & 5.11 & 3.36 \\
\bottomrule
\end{tabular}
\par\vspace{4pt}\noindent
\begin{minipage}{\columnwidth}
\scriptsize $Q_0=262{,}144$ fidelity evaluations, each using 8192 shots. $\dagger$ Diagnostic using exact products with the same probes and depth as SPSA-SLQ; it has additional derivative access and is not an equal-resource competitor or an error lower bound.
\end{minipage}
\end{table}

For $d\in\{32,64,128\}$, put $n_q=d/2$. Each qubit has angles $\alpha_j,\beta_j$, ordered as $\bm\varphi=(\alpha_1,\beta_1,\ldots,\alpha_{n_q},\beta_{n_q})^\top$. Parameter displacements $\bm\theta\in\R^d$ give
\begin{equation}
\bm\varphi(\bm\theta)=\bm\varphi_0+\bm O_d\bm\theta,
\qquad \bm O_d^\top\bm O_d=\bm I_d.
\label{eq:quantum-coordinates}
\end{equation}
Here $\bm O_d$ is a normalized Walsh-Hadamard matrix with random column signs and row permutation. Independent nominal angles satisfy $\alpha_{0j}\sim\mathcal U(0.45,1.25)$ and $\beta_{0j}\sim\mathcal U(-\pi,\pi)$; these and $\bm O_d$ are fixed within each instance. For product states, the fidelity from the nominal parameter $\bm0$ is
\begin{equation}
\begin{aligned}
f_0(\bm x)=\prod_{j=1}^{n_q}\tfrac12\bigl[1&+\cos\alpha_{0j}\cos\alpha_j(\bm x)\\
&+\sin\alpha_{0j}\sin\alpha_j(\bm x)\cos(\beta_j(\bm x)-\beta_{0j})\bigr].
\end{aligned}
\label{eq:fidelity}
\end{equation}
This product permits simulation of $16$, $32$, and $64$ qubits without storing an exponentially large state vector.

The Fubini-Study metric $\bm G$ measures local distances between pure states. With the convention of \cite{gacon2021}, the QFI for one copy of the state satisfies
\begin{equation}
\bm F_Q(\bm0)=4\bm G(\bm0)=-2\nabla_{\bm x}^2f_0(\bm x)\big|_{\bm x=\bm0}.
\label{eq:qfi-hessian}
\end{equation}
Consequently we use the scalar loss and estimate the log-determinant specified by
\begin{align}
\Phi_{0,\lambda}(\bm x)&=2[1-f_0(\bm x)]+\tfrac\lambda2\|\bm x\|_2^2,\nonumber\\
\bm H_\lambda&=\bm F_Q(\bm0)+\lambda\bm I_d,
\qquad \mathcal L_\lambda=\log\det\bm H_\lambda.
\label{eq:quantum-target}
\end{align}
Gacon et al.\ use diagonal regularization with weight $0.1$ on the Fubini--Study metric \cite{gacon2021}. Here both methods use the fixed weight $\lambda=0.1$ on the QFI scale. This defines a coordinate-specific score, not an intrinsic volume or a guarantee of jointly attainable parameter precision \cite{xing2020}.

A qubit has QFI equal to $\operatorname{diag}(1,\sin^2\alpha_{0j})$ in its angle coordinates. Transformation by \eqref{eq:quantum-coordinates} gives the exact reference:
\begin{align}
\bm F_Q&=\bm O_d^\top\bm D\bm O_d,\quad
\bm D=\operatorname{diag}(1,\sin^2\alpha_{01},\ldots,1,\sin^2\alpha_{0n_q}),\nonumber\\
\mathcal L_\star&=n_q\log(1+\lambda)+\sum_{j=1}^{n_q}\log(\sin^2\alpha_{0j}+\lambda).
\label{eq:quantum-truth}
\end{align}
The non-diagonal, full-rank QFI has shifted condition numbers $3.17$--$3.80$. Both estimators use sampled fidelities; direct use of the known product structure would favor \eqref{eq:quantum-truth}.

A shot is a binary trial with success probability $f_0(\bm x)$. Each function evaluation uses $S$ shots and returns $\widehat f_0(\bm x)=K_x/S$, where
\begin{equation}
K_x\sim\operatorname{Binomial}(S,f_0(\bm x)),\qquad S=8192.
\label{eq:finite-shots}
\end{equation}
Equation~\eqref{eq:oracle} holds with $\widetilde\Phi=2(1-\widehat f_0)+\lambda\|\bm x\|_2^2/2$ and $\epsilon=2(f_0-\widehat f_0)$, where $\xi$ represents shot randomness. Thus $\E\epsilon=0$ and $\operatorname{Var}(\epsilon)=4f_0(1-f_0)/S\le1/S$. Counts at the four evaluation points are independent. The ridge product $\lambda\bm v$ is added exactly; shots affect measurement accuracy, not the single-copy matrix.

Other QFI estimators use parameter-shift or linear-combination-of-unitaries techniques \cite{gacon2021}, or overlap measurements with Stein's identity \cite{halla2025}. Our comparison uses the four-fidelity full-matrix estimator of \cite{gacon2021}, rather than an exhaustive set of QFI methods.

The full-matrix baseline uses the asymmetric four-fidelity stencil in Eqs.\ (13) and (14) of \cite{gacon2021}, with two independent Rademacher perturbations and step $c_{\rm G}$. Denote its metric estimate by $\widehat{\bm G}_b$. Identity initialization and recursive averaging give $\overline{\bm G}_B=(\bm I_d+\sum_{b=1}^{B}\widehat{\bm G}_b)/(B+1)$. The reported estimate is
\begin{equation}
\widehat{\mathcal L}_{\rm G}=\log\det(4|\overline{\bm G}_B|+\lambda\bm I_d).
\label{eq:gacon-logdet}
\end{equation}
The factor four converts the metric to QFI; the ridge is added after spectral correction. Table~\ref{tab:quantum} calls this Full-QFI. SPSA-SLQ uses \eqref{eq:estimator} without spectral correction.

We compare $Q=Q_0,2Q_0,4Q_0$ queries, with $Q_0=262{,}144$ and $B=Q/4$. At $Q_0$, pooled mean squared relative error on three separate calibration instances, with two repetitions per instance and dimension, selects one setting per method. Both methods test eight candidates with equal calibration budgets, excluded from $Q$. Full-QFI uses $c_{\rm G}=0.075/\sqrt d$; SPSA-SLQ uses $c=0.20/\sqrt d$, $h=0.10$, $m=2$, and $(N,n_v)=(1024,32),(2048,32),(2048,64)$ at the respective budgets. Thus the first doubling increases HVP averaging and the second probe averaging. Global optimality is not claimed.

All 540 primary attempts in Table~\ref{tab:quantum} complete their assigned budgets. Full-QFI has smaller rRMSE at $d=32$; SPSA-SLQ does at $d=64,128$ for every budget. At $d=128$, $Q=4Q_0$, their RMSEs are $13.671$ [$13.340,13.991$] and $1.336$ [$1.047,1.587$]. These 95\% percentile-bootstrap intervals use 20,000 within-instance resamples and equal instance weights.

Full-QFI error at $d=128$ is not monotone in budget. Doubling the budget also fails to reduce RMSE uniformly in an original-settings control; details accompany the code.

Exact-HVP SLQ retains quadrature and probe errors. An independent exact-fidelity diagnostic (without shot noise) gives SPSA-SLQ RMSE $1.334$ at $d=128$, $Q=4Q_0$, over 15 runs (five per instance). SPSA variation and finite-difference bias remain; errors need not add in quadrature.

Shot variance decreases as $1/S$, with no added device noise. Neither exact-HVP error nor fixed-probe trace RMS is an irreducible floor. Settings, calibration, and diagnostics accompany the code.

\subsection{Example 2: High-Dimensional Fisher Information}
\label{subsec:example2}

We use the Gaussian signal-plus-noise family of \cite{spall2005}, related to state-space parameter estimation \cite{shumway1981}, to estimate information for a large joint mean and covariance parameter.

\begin{table}[t]
\centering
\caption{Example 2: three repetitions per method and dimension; listed evaluations are nominal per-run budgets. SP-FIM is the spectrally corrected full-FIM baseline \cite{spall2005}. Memory is peak allocated GPU tensor memory.}
\label{tab:example2}
\small
\setlength{\tabcolsep}{3pt}
\begin{tabular}{@{}lrrrr@{}}
\toprule
Method & RMSE & \shortstack{rRMSE\\(\%)} & \shortstack{Time\\(s)} & \shortstack{Memory\\(MiB)} \\
\midrule
\multicolumn{5}{@{}l}{$d=209$, $q=19$; 83,600 likelihood evaluations} \\
SP-FIM & 40.543 & 12.942 & 1.03 & 51.4 \\
SPSA-SLQ & 4.676 & 1.493 & 1.11 & 48.2 \\
\addlinespace
\multicolumn{5}{@{}l}{$d=989$, $q=43$; 395,600 likelihood evaluations} \\
SP-FIM & 1112.667 & 75.619 & 17.63 & 114.9 \\
SPSA-SLQ & 7.179 & 0.488 & 17.21 & 90.4 \\
\addlinespace
\multicolumn{5}{@{}l}{$d=3002$, $q=76$; 1,200,800 likelihood evaluations} \\
SP-FIM & 5691.811 & 127.804 & 140.09 & 360.9 \\
SPSA-SLQ & 41.290 & 0.927 & 129.39 & 126.5 \\
\bottomrule
\end{tabular}
\end{table}

We consider independent observations
\begin{equation}
\bm z_i\sim\mathcal N(\bm\mu,\bm\Sigma+\bm P_i),\qquad i=1,\ldots,n.
\end{equation}
Each $\bm z_i\in\R^q$. The unknown parameter is $\bm\theta=[\bm\mu^\top,\operatorname{vech}(\bm\Sigma)^\top]^\top$, with $\bm\Sigma\succ\bm0$ and $d=q+q(q+1)/2$. The operator $\operatorname{vech}$ lists lower-triangular covariance entries, including the diagonal, without scaling off-diagonal entries. The fixed nominal parameter $\bm\theta_0$ and the known noise covariances are specified by:
\begin{equation}
\bm\mu=\bm0,\quad\bm\Sigma=\bm I_q,\quad\bm P_i=\tfrac12\sqrt i\,\bm I_q,\quad n=30.
\label{eq:example2-setting}
\end{equation}
It is a specialization, not a reproduction of the particular $q=4$ realization in \cite{spall2005}. Put $a_i=1+\frac12\sqrt i$, $s_1=\sum_{i=1}^{n}a_i^{-1}$, and $s_2=\sum_{i=1}^{n}a_i^{-2}$. The exact FIM is diagonal with $q$ mean entries $s_1$, $q$ diagonal-covariance entries $s_2/2$, and $q(q-1)/2$ off-diagonal-covariance entries $s_2$. Therefore
\begin{equation}
\mathcal L_\star=q\log s_1+\frac{q(q+1)}2\log s_2-q\log2.
\label{eq:example2-truth}
\end{equation}
The analytical matrix supplies the reference and exact-HVP diagnostic; both estimators use likelihood values in this comparison.

In \eqref{eq:oracle}, take $\xi=\bm Z_k$, $\widetilde\Phi(\bm\theta;\bm Z_k)=-\log p(\bm Z_k\,|\,\bm\theta)$, and $\Phi=\E_{\bm\theta_0}\widetilde\Phi$. Each replication draws independent $\bm Z_k$ at \eqref{eq:example2-setting} and shares it across four perturbed evaluations. Given $\bm Z_k$, the loss is exact up to parameter-independent constants; $\epsilon=\widetilde\Phi-\Phi$ is sampling error, without injected noise.

The baseline follows the likelihood-only Hessian construction in Eqs.\ (3.1) and (3.2) of \cite{spall2005}, averaging one estimate per pseudodata set with two independent perturbations and $c_{\rm SP}=10^{-4}$. For the symmetric estimate $\widehat{\bm F}_{\rm SP}$, optional spectral correction gives $\widehat{\mathcal L}_{\rm SP}=\log\det|\widehat{\bm F}_{\rm SP}|$. Table~\ref{tab:example2} calls this SP-FIM; no eigenvalue floor is applied.

SPSA-SLQ uses maximum depth $m=2$, $n_v=1$, $c=h=5\times10^{-3}$, and $N=50d$; the baseline uses $B=2N$, giving each a nominal budget of $4B=400d$ likelihood evaluations per run. Runs use double-precision PyTorch on an NVIDIA GeForce RTX 3080 Ti. Table~\ref{tab:example2} used an earlier SPSA-SLQ variant: on a small residual it evaluates the smaller projection already constructed; Algorithm~\ref{alg:estimator} instead stops without an estimate. Actual depths and counts were not recorded, so listed budgets are upper bounds.

This diagonal FIM $\bm F$ satisfies $\bm z^\top\log(\bm F)\bm z=\log\det\bm F$ for every Rademacher probe, since $z_i^2=1$. Thus the exact trace-sampling term is zero; HVP and quadrature errors remain. Its three distinct eigenvalues favor quadrature; two-step exact-HVP relative errors are $0.472\%$, $0.241\%$, and $0.144\%$. These results do not imply that $m,n_v$ can remain fixed for general matrices or that $N=50d$ gives linear runtime.

At $d=3002$, SPSA-SLQ uses 65\% less peak GPU memory and 7.6\% less time (Table~\ref{tab:example2}). Synchronized GPU timings include pseudodata generation, likelihoods, and estimation; memory includes tensor batches. Likelihood $q\times q$ factorizations limit time savings.

\subsection{Example 3: Information-Based Sensor Placement}
\label{sec:example3}
On a synthetic planar monitoring region $[0,1]^2$, sensors record zero-mean fluctuations whose variance is modeled as a function of location. We compare equal-size layouts to reduce uncertainty in spatial log-variance coefficients \cite{joshi2009}. This independent-observation cosine-basis model is motivated by \cite{boukouvalas2014}, not a reproduction of its correlated Gaussian-process example. At $\bm s_i$, a sensor records $r=4$ independent fluctuations $Y_{it}\sim\mathcal N(0,\exp(\bm a(\bm s_i)^\top\bm\theta))$. The $d=p^2$ coefficients use $a_{uv}(\bm s)=c_u(s_x)c_v(s_y)/\sqrt d$, $0\le u,v<p$, with $c_0(x)=1$ and $c_u(x)=\sqrt2\cos(\pi ux)$ for $u>0$. Sensor observations are mutually independent.

For layout $\mathcal S$, $\bm A_{\mathcal S}$ contains rows $\bm a(\bm s_i)^\top$. At the fixed nominal parameter $\bm\theta_0=\bm0$, the random statistic $G_i=r^{-1}\sum_tY_{it}^2$ satisfies $rG_i\sim\chi_r^2$, with mean one and variance $2/r$ for $G_i$. Here $\eta_i=\bm a(\bm s_i)^\top\bm\theta$ is deterministic for a fixed layout and candidate parameter; $\bm G$ supplies the randomness. The centered negative log-likelihood and FIM are
\begin{align}
\widetilde\Phi(\bm\theta;\bm G)&=\frac r2\sum_i\left[\eta_i+G_i(e^{-\eta_i}-1)\right],\quad \bm\eta=\bm A_{\mathcal S}\bm\theta,\nonumber\\
\bm F_{\mathcal S}&=\left.\nabla_{\bm\theta}^2\E_{\bm\theta_0}\widetilde\Phi(\bm\theta;\bm G)\right|_{\bm\theta=\bm0}=\frac r2\bm A_{\mathcal S}^\top\bm A_{\mathcal S}.
\label{eq:sensor-fim}
\end{align}
In \eqref{eq:oracle}, $\xi=\bm G$ and $\epsilon(\bm\theta;\bm G)=(r/2)\sum_i(G_i-1)(e^{-\eta_i}-1)$. Replacing $G_i$ by its mean one gives $\Phi(\bm\theta)=(r/2)\sum_i[\eta_i+e^{-\eta_i}-1]$. Full column rank of $\bm A_{\mathcal S}$ ensures $\bm F_{\mathcal S}\succ\bm0$. We estimate $\mathcal L(\mathcal S)=\log\det\bm F_{\mathcal S}$ from scalar likelihoods; the SP methods redraw $\bm G$ per replication and share it across four evaluations. Layouts use the same basis and parameter units. Direct analytic evaluation is preferable here.

Set $p=12$ ($d=144$). Three fixed families each contain six 288-sensor layouts: two sensors per $12\times12$ cell, with independent offsets uniform on $[-0.325,0.325]^2/12$ from cell centers. Thirty repetitions yield 540 estimates and 90 selections per method; reference condition numbers are $6.95$--$13.76$. Each whole-likelihood method uses $Q=8d^2=165{,}888$ evaluations per layout. FD performs four sweeps, reusing one nominal dataset for $2d^2$ evaluations each since $\widetilde\Phi(\bm0;\bm G)=0$. Full SP, blockwise feedback, and grouped independent-perturbation (IP) methods adapt \cite{spall2008acc}; feedback freezes the estimate within each block. Full-matrix outputs use $\sum_i\log|\lambda_i(\widehat{\bm F})|$ without requiring positive definiteness. SP steps are $0.1/\sqrt d$; SPSA-SLQ directional and coordinate FD steps are $0.1$.

Separate calibration against analytic truth fixes 16 feedback blocks, 8 IP groups, and SPSA-SLQ settings $m=2$, $n_v=16$, $N=1296$. Tuning costs are excluded from $Q$ and unmatched. Here $Q$ measures likelihood computation, not data acquisition: FD uses 4 simulated datasets per run and layout, versus 41,472 for SPSA-SLQ. No physical observations are collected. Layouts share random numbers within each method.

\begin{table}[t]
\centering
\caption{Example 3 at $d=144$: 18 layouts and 90 selections. RMSE is for the estimated $\log\det\bm F_{\mathcal S}$; brackets show $95\%$ bootstrap intervals. The best point estimate in each column is bold.}
\label{tab:sensor-placement}
\setlength{\tabcolsep}{3pt}
\footnotesize
\begin{tabular}{lccc}
\toprule
Method & \shortstack{Log-det RMSE $\downarrow$\\{[interval]}} & \shortstack{Best (\%)\\$\uparrow$} & \shortstack{Regret\\$\downarrow$} \\
\midrule
Full SP & 37.51 [37.25, 37.79] & 35.6 & 1.20 \\
Feedback SP & 21.59 [21.20, 21.96] & 32.2 & 1.78 \\
Grouped IP SP$^{\dagger}$ & 32.85 [32.59, 33.13] & 41.1 & 0.67 \\
Coordinate FD & 5.50 [4.94, 6.04] & \textbf{77.8} & \textbf{0.05} \\
SPSA-SLQ & \textbf{2.48} [2.28, 2.68] & 54.4 & 0.37 \\
\bottomrule
\end{tabular}
\par\vspace{4pt}\noindent
\begin{minipage}{\columnwidth}
\scriptsize Best: frequency of choosing the true best of six layouts. Regret: loss in true log-determinant from that choice. $\downarrow$/$\uparrow$: smaller/larger is better. $\dagger$ Grouped IP accesses 8 component likelihoods: 8$Q$ scalar values with the same number of sensor-likelihood terms as $Q$ whole-likelihood values.
\end{minipage}
\end{table}

Each repetition selects the largest estimated log-determinant among six layouts. For layout selection, regret is primary: the mean of $\max_{\mathcal S}\mathcal L(\mathcal S)-\mathcal L(\widehat{\mathcal S})$ over 90 selections, where $\widehat{\mathcal S}$ is the chosen layout. Best measures exact recovery; log-det RMSE measures estimation accuracy (Table~\ref{tab:sensor-placement}). Within-family bootstrap resampling preserves all six paired layouts per repetition. FD gives higher Best and smaller regret despite SPSA-SLQ's lower log-det RMSE. Pairwise log-determinant difference RMSEs are $0.80$ for FD and $2.59$ for SPSA-SLQ, pooling 15 pairs per selection. Common score shifts cancel, so lower log-det RMSE need not improve selection.

\section{Conclusion}
\label{sec:conclusion}

SPSA-SLQ estimates positive-definite Hessian log-determinants from possibly noisy function values without forming the matrix. It covers fixed-data and expected criteria, including Fisher information and posterior curvature.

The 3002-parameter example gives lower log-determinant error and 65\% less peak GPU memory than the basic full-matrix estimator. The sensor study gives lower log-determinant RMSE than the compared full-matrix methods, although coordinate finite differences make better layout choices. SPSA-SLQ's accuracy and cost depend on the spectrum, noise, and algorithm settings. The method provides a way to carry out log-determinant calculations for uncertainty assessment and experimental design in system identification and control when direct derivatives are difficult to obtain.

\bibliographystyle{ieeetr}
\bibliography{references_ACC_v15}
\end{document}